# Spectral Oscillations in Backscattering of Light from a Biological Cell

Alexander Heifetz[1], Alexander Patashinski[2], Vadim Backman[3]

[1]*Nuclear Engineering Division, Argonne National Laboratory, Argonne, IL 60439*

[2]*Department of Chemistry, Northwestern University, Evanston, IL 60208*

[3]*Department of Biomedical Engineering, Northwestern University, Evanston, IL 60208*

Possible origins of the experimentally observed beat-like signal in elastic light backscattering from an epithelial biological cell are identified. The source of the beat-like signal is most likely the cellular nucleus, which is a spheroidal particle with an optically sharp edge, low relative index of refraction, and a size parameter in the resonant range; another possible source is an agglomeration of cell organelles of a similar size. Using Mie theory computer simulations for a dielectric sphere with biological nucleus-like parameters, we have shown that backscattering as a function of size parameter has a beat-like pattern. The high frequency periodic oscillation is the single scattering contribution, while multiple scattering events give slowly varying oscillatory contributions. With increasing contrast, the backscattering signal changes from a pure sinusoid to that of a beat signal. The beat-like pattern in scattering from a low contrast particle does not depend on any particular features of the sphere; a similar pattern is observable in scattering from star-like non-spherical objects with a sharp optical contrast and smooth shape; in this case, the high frequency of oscillations depends on the object orientation relative to incident light direction.

Recently, there has been a significant interest in studying biological cell structures with visible light elastic scattering [1,2]. In elastic backscattering spectroscopy, a cell is illuminated by a polychromatic plane wave, and normalized backscattered intensity in the far field is recorded as a function of the wavelength. Under visible light illumination, many cell types, such as the epithelial cells, where most human carcinomas arise, can be very well approximated as lossless dielectrics. Experimental data indicates that elastic backscattering of visible light from epithelial cells exhibits a beat-like pattern, with fast spectral oscillations modulated by a slow spectral envelope [1].

Assuming that light absorption and nonlinear effects are negligible, one can relate the spectral dependence of backscattered intensity to morphology (size, shape and optical index of refraction) of the scatterer. Spectral partial wave resonances in backscattering from a lossless dielectric sphere of high enough optical contrast are predicted by Mie theory [3-5]. There are many partial waves giving a substantial contribution to scattering from a sphere of a radius larger than the wavelength. In the limit of very low contrast, iterations of Maxwell equations can be used; the first iteration represents single scattering, while higher iterations describe multiple scattering.

In this letter, we show that dielectric particles with biologically relevant parameters of a low refractive index contrast and sizes on the order of ten light wavelengths lay between the two asymptotic ranges, the Mie resonances and single scattering. For such dielectric particles, Mie resonances are very wide: each partial wave gives a smooth spectral dependence. At the same time, single scattering is not the only significant contribution to the observed signal . To see the relevance of different approximations and contributions of different scattering mechanism, we calculated backscattering as function of the size parameter for same-size dielectric sphers with different optical contrast levels. It appears that for biological cell parameters, single scattering from a sphere or spheroid produces fast periodic oscillations modulated by multiple order scattering contributions having smoother periodic frequency dependence. These results are readily extendable to arbitrary biological shapes because single-scattering contribution can be easily calculated for scatterers lacking spherical symmetry.

Born series for backscattering from uniform spheres has been studied in [6-8]. One can treat the scattered signal as an interference of two terms: the amplitude of the first order Born approximation (single scattering) and the combined amplitude of the second and all higher Born approximation orders (multiple scattering). The amplitude of multiple scattering was calculated for quantum scattering in [6,7], and for optical scattering in [8] under the assumption of single hard scattering event that turns the k-vector around. The forward scattering events before and after the hard scattering event were treated in the eikonal approximation. The resulting far-field backscattering signal as a function of the size parameter has a beat-like pattern, where low frequency oscillation is introduced by multiple scattering orders. Another way to see the contributions of single and multiple scattering is to calculate, using Mie theory, the scattered intensity for gradually increasing optical density of a fixed size sphere, starting from very low optical density values.

Consider a hypothetical light scattering experiment, where an isolated cell suspended in a physiological buffer (such as Phosphate Buffer Solution or PBS) is illuminated by a plane wave. A biological cell contains a distribution of scatterers with sizes ranging from nanometers (ribosome) to microns (nucleus). The refractive index $m_0$=1.36 of the physiological buffer is matched with that of cellular cytoplasm, so that only optical heterogeneities inside the cell scatter light. The nucleus of a typical epithelial cell is a spheroidal structure with an average radius $a \sim 5\mu m$, and has a sharp boundary. Average refractive index values for an epithelial cell nucleus quoted in the literature are in the range from $m_n$=1.38 to $m_n$=1.46 [2]. Assuming the median value, the average refractive index contrast of the nucleus (relative to the cytoplasm and the physiological buffer) is $m \equiv m_n/m_0 \sim 1.05$. For visible light illumination (400nm$\leq \lambda_0 \leq$700nm in free space and 300nm$\leq \lambda \leq$500nm in the physiological buffer), the size of the nucleus is many times the incident wavelength. Single scattering from a spheroid approximating the nucleus exhibits fast spectral oscillations, so it is reasonable to assume that the dominant signal of the oscillatory backscattering spectrum is due to the nucleus. Randomly positioned organelles are much smaller than the wavelength of the incident light, and contribute to a featureless spectral background curve characteristic of Rayleigh scattering. However, small organelles may agglomerate into larger clusters having an effective diameter substantially larger than the wavelength of the incident light. In this case, these agglomerates may also contribute to the oscillatory spectrum.

We used computer simulations for the backscattering from dielectric spheres; the radius of the spheres was fixed at $a$=5μm, and the range of optical densities scanned was chose from values much smaller to much larger than that for a nucleus. MATLAB codes for the Mie series solution written by C. Mätzler [9] have been used. The non-dimensional size parameter $x = ka$ was in the range of $x \sim 100$, where $k = 2\pi/\lambda$ is the wavenumber in the background medium. The backscattering cross-section efficiency $Q_b$ (backscattering cross-section normalized by the area of the particle) is given as an the sum over angular momentum number $n$ [4,9]

$$Q_b(x) = x^{-2} \left| \sum_{n=1}^{\infty} (2n+1)(-1)^n (a_n - b_n) \right|^2, \quad (1)$$

where

$$a_n = \frac{m^2 j_n(mx)[xj_n(x)]' - j_n(x)[mxj_n(mx)]'}{m^2 j_n(mx)[xh_n^{(1)}(x)]' - h_n^{(1)}(x)[mxj_n(mx)]'},$$

$$b_n = \frac{j_n(mx)[xj_n(x)]' - j_n(x)[mxj_n(mx)]'}{j_n(mx)[xh_n^{(1)}(x)]' - h_n^{(1)}(x)[mxj_n(mx)]'} \qquad (2)$$

$j_n(z)$ is the spherical Bessel function, $h_n^{(1)}(z) = j_n(z) + i y_n(z)$ is the spherical Hankel function of the first kind, and $y_n(z)$ is the spherical Neumann function [10]. The primes are derivatives with respect to the argument $z = x$ or $mx$. Note that the sum is converging for large values of $n$ and can be truncated; it has been estimated that $n_{max} = 4x^{1/3} + x + 2$ [11]. The details of computer implementation of these series can be found in [9].

The backscattering cross-section efficiency $Q_b$ calculated from Mie theory is plotted in Figures 1(a)-(d) as function of $x$ for different values of $m$-1. Fig. 1(a) shows oscillations in the backscattering cross-section for $m$-1 = 0.001 or $x(m$-1$) \sim 0.1$. Note that $Q_b$ can be expanded in power series in $x(m$-1$)$,, and for $x(m$-1$)<<1$ the power series is well approximated by the first order term representing single scattering. Then, the Mie formula for the backscattering efficiency $Q_b$ (Eq. 1) reduces to that of the first Born approximation for scattering on a sphere

$$Q_b(x) = (m-1)^2 [\cos(2x) - \sin(2x)/(2x)]^2 \qquad (3)$$

As a test of accuracy, we have verified numerically that for $m$-1< 0.001 (see Fig. 1(a)) $Q_b$ calculated with Mie theory is accurately given by Eq. (3). In this range of optical parameters, single scattering clearly dominates.

For optical density values above $m$-1 =0.001, the behavior of $Q_b$ deviates from that given by Eq. (3); deviation increases with rising optical contrast, indicating a more pronounced contribution of terms in the scattering amplitude that are of second and higher orders in $x(m$-1$)$.

The graph in Figure 1(b) for $m-1 = 0.01$, which is lower than the value for a cellular nucleus, shows that single scattering is still the dominant contribution to the measured backscattering signal. However, corrections due to multiple scattering are noticeable, and they result in intensity modulation and shifts in the positions of minima and maxima. Computer simulations data for $m-1 = 0.05$ are shown in Fig. 1(c). The parameters for this case are most closely related to that of a biological nucleus, and there is a clearly pronounced multiple scattering contribution to the overall signal. In figure 1(d), where $m-1 = 0.1$, a value above that of the nucleus, multiple scattering dominates the backscattering spectrum, and one can see a wideband envelope spectral structure. For values of $m-1 > 0.2$, significantly higher than the optical contrast typical for cellular structures, Mie resonances become detectable, first as very wide spectral peaks. Such values of the optical density are characteristic of polystyrene beads in water, which are often used as tissue phantoms. For values higher than $m-1 > 0.3$, the spectral peaks rapidly increase and become narrow resonances that are orders of magnitude higher than the single scattering oscillations. Such values of relative refractive index are typical for water droplets or dielectric microspheres in air.

Using computer simulations, we tested that, for any given case of such resonances, they are contained in only one partial wave amplitude of the corresponding expansion. The optical parameters of a scatterer producing these narrow spectral peaks are outside the range of cellular structure parameters. It is known that a small deviation of the scatterer shape from perfect spherical or cylindrical symmetry leads to a substantial suppression of Mie resonances [12]. In contrast to that, the single scattering spectral oscillating signal is not sensitive to small changes of the scatterer geometry. Note that in the perturbation theory [13], every $n$-th term in the power series expansion of Mie cross-section gives higher order corrections to the coefficient in the first order term describing high frequency spectral oscillation; single scattering represents the lowest order term of this coefficient.

The interference of the first order scattering amplitude with that of the multiple order scattering explains why the period of the high frequency oscillations in Figures 1(a)-(d) slightly decreases with increasing value of *m*-1. Both single scattering and multiple scattering amplitudes are characterized by their absolute values and phases, and their phase difference is a slow-varying function of the wave vector. One can write the complex backscattering amplitude containing all orders of scattering $F(x) = F_s(x) + F_m(x)$ as a sum of a single scattering term $F_s = |Q_{b,s}(x)|^{1/2} \exp(i\varphi_s(x))$, and a term due to the second and all higher order scattering events $F_m = |Q_{b,m}(x)|^{1/2} \exp(i\varphi_m(x))$, where $Q_{b,s}(x)$, $\varphi_s(x)$ and $Q_{b,m}(x)$, $\varphi_m(x)$ are the scattering efficiencies and phases of the single scattering and multiple scattering contributions. A similar approach was taken in [6]. Hence, the backscattering efficiency $Q_b(x)$ due to all scattering orders is

$$Q_b(x) = |F(x)|^2 = Q_{b,s}(x) + Q_{m,s}(x) + 2|Q_{b,s}(x)Q_{m,s}(x)|^{1/2} \cos[\varphi_m(x) - \varphi_s(x)] \quad (4)$$

As can bee seen from Eq. (4), modulation of the fast oscillatory part by the slowly varying part shifts the positions of minima and maxima, and thus changes the values of the backscattering efficiency $Q_b$ at these points. These effects are clearly illustrated by Figs. 1 (c) and 1(d). The quantities $Q_{b,s}(x)$, $Q_{b,m}(x)$, and the phase difference $\phi_m(x) - \phi_s(x)$ can be extracted from experimental data from the best fit of the experimental data to Eq. (4).

Unlike the case of Mie resonances, the beat-like oscillatory signal in scattering from a low contrast particle is not an exclusive feature of perfect spherical geometric shapes. Rather, spectral oscillations observed in first order scattering are due to the sharp edge of the particle, and disappear when the transition at the boundary between two media becomes smooth on the length-scale of the optical wavelength. The single scattering signal is proportional to the Fourier harmonic $I(K)$ of the scatterer optical density [14]

$$I(K) = \int U(\mathbf{r})\exp(-i\mathbf{K}\cdot\mathbf{r})dV, \tag{5}$$

where $U(\mathbf{r}) = m^2(\mathbf{r}) - 1$ is the scattering potential, and the scattering vector amplitude is $K = 2k\sin(\theta/2)$, where $\theta$ is the angle between the directions of the incident and scattered waves. Choosing the z-axis to be in the direction of the vector $\mathbf{K}$, one can rewrite (5) as

$$I(K) = \int S(z)e^{-iKz}dz, \quad S(z) = \int U(x,y,z)dxdy. \tag{6}$$

The function $S(z)$ is the projection of the optical density on the z-axis: for a given $z$, integration in Eq. (6) is carried out over all $x$ and $y$ inside the volume of the scatterer. For a finite-size scatterer $S(z)$ vanishes at infinity. For the special case of a homogeneous sphere with radius $a$ and constant relative refractive index $m$, $S(z)=2\pi(m^2-1)(a^2-z^2)$ for $|z|<a$, and $S(z)=0$ otherwise. For an object with a sharp boundary, $S(z)$ is singular at projection points $z_i$ of scatterer edges onto the z-axis; near $z_i$, the singular part of $S(z)$ can be represented as (we assume that $S(z)=0$ for $z > z_i$)

$$S(z)|_i = H(z-z_i)\sum_{k=1}^{\infty}\frac{s_i^{(k)}}{k!}(z-z_i)^k, \quad s_i^{(k)} = \frac{d^k S(z)}{dz^k}|_{z\to z_i+0}, \tag{7}$$

where $H(x)$ is the Heaviside step function: $H(x)=0$ for $x<0$, and $H(x)=1$ for $x>0$. The contribution of the singularity in Eq. (6) to $I(K)$ is an oscillating function

$$I(K)|_i = \sum_{n=1}^{\infty}\frac{s_i^{(n)}}{n!}\int H(z-z_i)(z-z_i)^n e^{-iKz}dz = e^{-iKz_i}F_i(K),$$

$$F_i(K) = \sum_{n=1}^{\infty}\frac{s_i^{(n)}}{n!}\int H(z)z^n e^{-iKz}dz = e^{-iKz_i}\sum_{n=1}^{\infty}s_i^{(n)}\left(\frac{1}{iK}\right)^{n+1}. \tag{8}$$

For a dielectric sphere of radius *a*, there are two singular projection points $z_1=a$, $z_2=-a$, and $s_i^{(n)}$ differ from zero only for $n=1$ and $n=2$; hence the resulting formula for scattering intensity coincides with Eq. (3). For a non-spherical object with a sharp boundary, the positions of singular points, and thus the period of spectral oscillations for both first order and multiple order scattering events depend on orientation of the particle relative to scattering vector ***K***.

For an object with smooth boundaries, the step at $z_i$ is replaced by a continuous change in an interval $\Delta z_i$. When this interval is small ($K\Delta z_i<<1$), Eq. (7) may be used as a valid approximation; for a gradual transition boundary, the spectral oscillations become washed out.

In summary, we have identified possible origins of the experimentally observed beat-like spectral signal in elastic light backscattering from an epithelial biological cell: the most likely source of this signal is the cellular nucleus having a spheroidal shape with an optically sharp edge, a low relative index of refraction, and a size parameter in the resonant range. Using Mie theory computer simulations, we have tested that a beat-like backscattering spectrum is obtained in scattering from a dielectric sphere with biological nucleus-like parameters; oscillation with a small spectral period appears due to single scattering at the singular points, while the spectrally smooth envelope is due to multiple order scattering events. The backscattering signal changes with increasing contrast from a pure spectral sinusoid to that of a beat signal. The beat-like spectral pattern in scattering from a low contrast particle is a feature of light scattering from smooth-shaped objects with a sharp optical edge.

For the part of this work done at Northwestern University, Alexander Heifetz was supported by American Cancer Society / Canary Foundation Postdoctoral Fellowship in Early Detection. A.H. would like to thank N. Sami Gopalsami and A.C. Paul Raptis for the access to facilites at the Argonne National Laboratory.

**Appendix**

To see the relations of single scattering and the partial amplitudes expansion, one takes into account that in Mie theory, the backscattering cross-section efficiency is given by [3-5]

$$Q_b(x) = x^{-2} \left| \sum_{n=1}^{\infty} (2n+1)(-1)^n (a_n - b_n) \right|^2 \tag{A1}$$

$$a_n = \frac{m^2 j_n(mx)[xj_n(x)]' - j_n(x)[mxj_n(mx)]'}{m^2 j_n(mx)[xh_n^{(1)}(x)]' - h_n^{(1)}(x)[mxj_n(mx)]'} \tag{A2a}$$

$$b_n = \frac{j_n(mx)[xj_n(x)]' - j_n(x)[mxj_n(mx)]'}{j_n(mx)[xh_n^{(1)}(x)]' - h_n^{(1)}(x)[mxj_n(mx)]'} \tag{A2b}$$

where $n$ is the angular momentum number, $j_n(z)$ is the spherical Bessel function and $h_n^{(1)}(z)$ is the spherical Hankel function of the first kind, which is $h_n^{(1)}(z) = j_n(z) + iy_n(z)$, where $y_n(z)$ is the spherical Neumann function. The primes are derivatives with respect to the argument $z = x$ or $mx$.

In Equation (A1), Mie backscattering cross-section is given as an infinite sum over the angular momentum number. One expands the coefficients $a_n$ and $b_n$ in power series in the non-dimensional parameter $(m-1)x$, which characterizes the strength of the scattering potential [3]. The first order of this expansion has to yield the same expression for Mie backscattering cross-section efficiency as the one obtained from the first Born approximation for a sphere [3-5]. To expand the coefficients in powers series in $(m-1)x$, $a_n$ and $b_n$ can be written as

$$a_n = \frac{m^2 a - b}{m^2 c - d} = \frac{(m^2-1)a + (a-b)}{(m^2-1)c + (c-d)} \tag{A3a}$$

$$b_n = \frac{a-b}{c-d} \tag{A3b}$$

where $a = j_n(mx)[xj_n(x)]'$, $b = j_n(x)[mxj_n(mx)]'$, $c = j_n(mx)[xh_n^{(1)}(x)]'$, $d = h_n^{(1)}(x)[mxj_n(mx)]'$,

where $m^2 - 1 = (m-1)(m+1) = \xi(\xi+2)$, and we have defined $\xi = m - 1$, Therefore, we can express $a_n$ in power series to obtain

$$a_n - b_n = \xi 2\frac{bc-ad}{(c-d)^2} + \xi^2\left[\frac{bc-ad}{(c-d)^2} - 4\frac{bc-ad}{(c-d)^2}\frac{c}{c-d}\right] + O(\xi^3) \tag{A4}$$

Note that in backscattering the zero-order term has been cancelled out. To obtain the first order term, we evaluate the fraction by setting $m=1$. Using the recursion relation

$$[zf_n(z)]' = zf_{n-1}(z) - nf_n(z), \tag{A5}$$

where $f_n(z)$ can be either $j_n(z)$ or $h_n^{(1)}(z)$ and the cross-product relation [6]

$$j_n(x)y_{n-1}(x) - j_{n-1}(x)y_n(x) = x^{-2} \tag{A6}$$

which can also be written as $j_n(x)h_{n-1}^{(1)}(x) - j_{n-1}(x)h_n^{(1)}(x) = ix^{-2}$, we obtain $\tag{A7}$

$$bc - ad = ix^{-1}j_n(x)[xj_n(x)]' \tag{A8a}$$

$$c - d = ix^{-1} \tag{A8b}$$

Therefore,

$$(bc-ad)/(c-d)^2 = -ix\left[j_n^2(x) + xj_n(x)j_n'(x)\right] \tag{A9}$$

In order to evaluate the infinite sum, we use the identity [10]

$$\sum_0^\infty (2n+1)(-1)^n j_n^2(x) = \sin(2x)/(2x) \tag{A10}$$

and its derivative

$$\sum_0^\infty (2n+1)(-1)^n 2j_n(x)j'_n(x) = [2x\cos(2x) - \sin(2x)]/(2x^2). \tag{A11}$$

Therefore

$$\sum_{1}^{\infty}(2n+1)(-1)^n(j_n^2(x)+xj_n(x)j_n'(x)) = \sum_{0}^{\infty}(2n+1)(-1)^n(j_n^2(x)+xj_n(x)j_n'(x)) - $$
$$-j_0^2(x) - xj_0(x)j_0'(x) = [\cos(2x) - \sin(2x)/(2x)]/2 \qquad (A12)$$

where we have used $j_0(x) = \sin x/x$ and $j_0'(x) = (x\cos x - \sin x)/x^2$ [10]. Therefore, the first order approximation of Mie backscattering efficiency is

$$Q_b(x) = (m-1)^2[\cos(2x) - \sin(2x)/(2x)]^2 \qquad (A13)$$

which is identical to the first Born (single scattering) approximation for scattering from a uniform sphere. For $x \gg 1$, backscattering can be reduced to

$$Q_b(x) = (m-1)^2 \cos^2(2x) \qquad (A14)$$

**References:**

1. L.T. Perelman, V. Backman, M. Wallace, G. Zonios, R. Manoharan,, A. Nusrat, S. Shields, M. Seiler, C. Lima, T. Hamano, I. Itzkan, J. Van Dam, J.M. Crawford, M.S. Feld, *"*Observation of periodic fine structure in reflectance from biological tissue: A new technique for measuring nuclear size distribution*"*, Phys. Rev. Lett., 80, 627 (1998).

2. L.T. Perelman, V. Backman, *Light Scattering Spectrosocopy of Epithelial Tissues: Principles and Applications*, in *Handbook of Optical Biomedical Diagnostics*, V.V. Tuchin, Ed., SPIE Press (2002).

3. H.C. van de Hulst, *Light Scattering by Small Particles*, Dover (1981).

4. C.F. Bohren, D.R. Huffman, *Absorption and Scattering of Light by Small Particles*, Wiley-VCH (2004).

5. A. Ishimaru, *Wave Propagation and Scattering in Random Media*, Wiley-IEEE Press (1999).

6. S.K. Sharma, D.J. Somerford, "An approximation method for the backward scattering of light by a soft spherical obstacle," J. Mod. Opt. 41, 1433-1444 (1994).

7. L.I. Schiff, Approximation method for high-energy potential scattering," Phys. Rev. 103, 443-453 (1956).

8. J.F. Reading and W.H. Bassichis, High energy scattering at backward angles," Phys. Rev. D 5, 2031-2041 (1972).

9. C. Mätzler, "MATLAB Functions for Mie Scattering and Absorption," University of Bern Research Report No. 2002-08 (2002).

10. M. Abramowitz, I.A. Stegun, *Handbook of Mathematical Functions: with Formulas, Graphs, and Mathematical Tables*, Dover (1965).

11. W.J. Wiscombe, "Improved Mie scattering algorithms," Appl. Opt. 19, 1505-1509 (1980).

12. P.W. Barber, R.K. Chang, Ed., *Optical Effects Associated with Small Particles, World Scientific* (1988).

13. L.D. Landau, L.M. Lifshitz, *Quantum Mechanics: Non-Relativistic Theory, Volume 3*, Butterworth-Heinemann, 3$^{rd}$ edition (1981).

14. M. Born and E. Wolf, Principles of Optics, 7$^{th}$ edition, Cambridge University Press (1999).

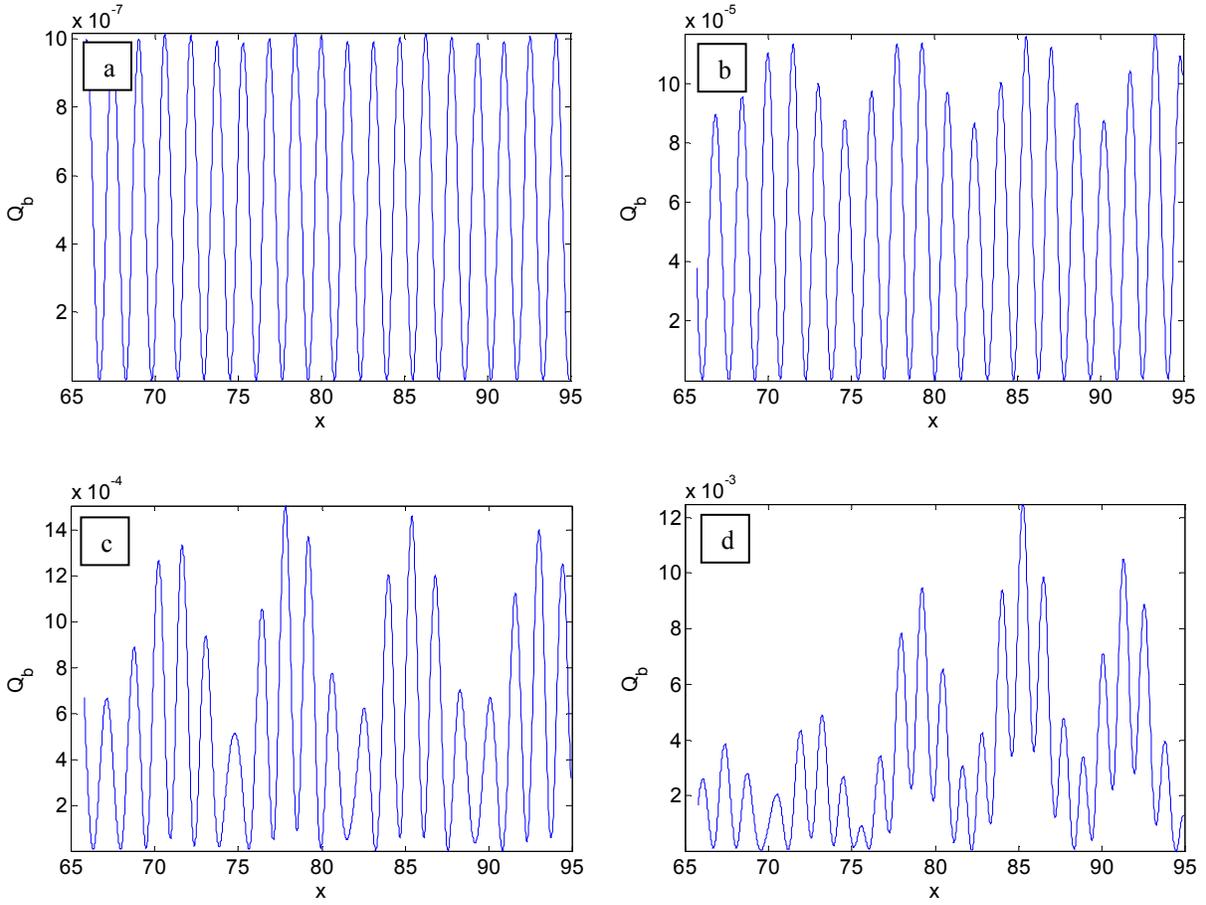

**Figure Captions**

Figure 1: Mie backscattering cross-section efficiency $Q_b$ as a function of the size parameter $x$ for a and (a) $m-1=0.001$, (b) $m-1=0.01$, (c) $m-1=0.05$, (d) $m-1=0.1$